\documentclass[12pt,onecolumn, draftclsnofoot]{IEEEtran}

\usepackage{cite}
\usepackage{graphicx}
\usepackage{psfrag}
\usepackage{subfigure}
\usepackage{url}
\usepackage{stfloats}
\usepackage{amsmath}
\usepackage{array}

\begin{document}

\title{L-MAC: Location-aware MAC Protocol for Wireless Sensor Networks}


\author{Jason Chen$^1$, Yang Xi$^2$\\ 
$^1$Samsung Research, USA\\
$^2$Quanzhou University of Information Engineering, China}


\maketitle

\begin{abstract}
This paper presents the design, implementation and performance evaluation of a location MAC protocol, called
L-MAC, for wireless sensor networks. L-MAC is a combination of TDMA and CSMA while offsetting the high overhead of time slot
assignment by allocating the time slots to sensor nodes based on their location information. This design avoids  high
computation complexity of time slot assignment incurred by node mobility and node failure. The area which the wireless
sensor network occupies is divided into blocks and each block is associated with an inter-block time slot and an intra-block
time slot. In the inter-block time slot, the sensor nodes stay active and receive the packets from nodes outside of the
block. In the intra-block time slot, the sensor nodes communicate with peer nodes in the same block under CSMA. Sensor nodes
stay sleep in all other time slots unless they have traffic to send. L-MAC is implemented and evaluated in NS-2.
\end{abstract}

\section{Introduction}

A radio channel is able to provide a certain amount of channel capacity if the access to the channel is well coordinated in
time, frequency, code and space domains. Medium access control (MAC) plays a key role in wireless sensor networks. A good
MAC protocol can improve the performance of wireless sensor networks in several aspects, such as channel utilization,
end-to-end delay, throughput and energy consumption.

A sensor node is extremely limited in power, computational capacities and memory. Due to these basic constraints, design of
MAC protocols is generally different to design of traditional MAC protocols. Energy consumption minimization becomes the
most important objective other than objectives such as throughput maximization, delay minimization and fairness. The major
sources of energy waste in wireless sensor networks include collision, overhearing, control overhead and idle listening.
When collision happens, a transmitted packet is corrupted and has to be discarded, and the following retransmissions
increase energy consumption. Overhearing refers to a node picks up packets that are destined to other nodes, while idle
listening refers to a node listens to receive possible traffic that does not exist.

Typical MAC protocols include time division multiple access (TDMA), frequency division multiple access (FDMA), code division
multiple access (CDMA), and contention-based protocols such as carrier sensing multiple access (CSMA). TDMA can avoid the
collision by scheduling transmission of different sensor nodes in different time slots. However, TDMA has several other
disadvantages. Firstly, algorithm for efficient time slot assignment is not trivial. It often requires a centralized node to
find a collision-free schedule. Furthermore, developing an efficient schedule with a high degree of channel utilization is
very hard. Secondly,  TDMA needs time synchronization. high-precision synchronization leads to high control overhead and
energy consumption. Thirdly, wireless sensor networks may experience frequency topology change because of time-varying
channel, node movement, node failure and physical environmental changes. The dynamic topology requires time slot assignment
to be updated in a timely manner, which also leads to high computation complexity and energy consumption. Lastly, channel
utilization by TDMA is low in case the traffic is low.

CSMA is a common MAC protocol in wireless networks. It becomes popular because of its simplicity, flexibility and
robustness. It does not require clock synchronization and global topology information. It handles the dynamic topology, such
as node joining and node failure, without extra operations. The disadvantage of CSMA is collision. Collision can happen in
any two-hop neighbors of a sensor node. While collision among one-hop neighbors can be greatly reduced by carrier sensing
before transmission, carrier sensing does not work beyond one hop. This problem is also call hidden terminal problem, which
can cause a serious throughput degradation especially in high data rate sensor applications. Although RTS/CTS mechanism can
alleviate the hidden terminal problem, it also introduces high control overhead.

In this paper, we present a new location-based MAC protocol, called L-MAC, for wireless sensor networks. L-MAC is a
combination of TDMA and CSMA while reducing the high overhead of time slot assignment by allocating the time slots to sensor
nodes based on their location information. This design avoids high computation complexity of time slot assignment incurred
by dynamic topology of wireless sensor networks, such as node mobility and node failure. In our design, the area of interest
is divided into blocks with equal size and each block is associated with an inter-block time slot and an intra-block time
slot. Sensor nodes obtain their block ID by comparing their location and the block coordinates, then obtain their time
slots. In inter-block time slot, the sensor nodes stay active and receive the packets from nodes outside of the block. In
the intra-block time slot, the sensor nodes communicate with peer nodes in the same block under CSMA mechanism. Sensor nodes
stay sleep in all other time slots unless they have traffic to send. Time slots are reused throughout the whole networks to
reduce the packet delay. The reuse rule is designed to minimize the inter-block interference by considering the inter-block
distance and sensor node transmission range.

The paper is organized as follows. Section \ref{rw} gives a brief literature survey on MAC protocol designs for wireless
sensor networks. Section \ref{pd} introduces details of L-MAC protocol design. The simulation results of L-MAC and
comparison to existing MACs will be given in section \ref{sr}. The conclusion is drawn in section \ref{cl}.

\section{Related Work} \label{rw}
MAC protocol design receives a lot of attentions in sensor network research community. Various MAC protocols
\cite{mac-s}-\cite{mac-d} were proposed in recent years.

S-MAC \cite{mac-s} introduces periodic listening and sleep mechanism to save the energy of sensor node. The listen time of
sensor node is reduced by going into periodic sleep mode. During sleep, the node turns off its radio, and sets a timer to
awake itself later. In order to reduce control overhead, neighbor nodes synchronize to each other so that they have the same
duty cycle after synchronization. The disadvantage is that the latency is increased due to the periodic sleep of each node.
Moreover, the delay can accumulate on each hop. The topologies of the experiments are five nodes forming two-hop networks
and ten nodes forming a straight line.

T-MAC \cite{mac-t} is an adaptive energy-efficient MAC protocol for wireless sensor networks. It manages to save more energy
by assigning the duty cycle of each node adaptively, other than fixed duty cycle in S-MAC.  The novel idea of the T-MAC
protocol is to reduce idle listening by transmitting all messages in bursts of variable length, and sleeping between bursts.
The length is determined dynamically and the active time is ended when hearing for nothing for a given time. The topology of
the experiments is 100 nodes forming a grid topology.

B-MAC \cite{mac-b} is a CSMA protocol for wireless sensor networks. It provides a flexible bidirectional interface to obtain
ultra low power operation, effective collision avoidance, and high channel utilization. To achieve low power operation,
B-MAC employs an adaptive preamble sampling scheme to reduce duty cycle and minimize idle listening. B-MAC also supports
on-the-fly reconfiguration and provides bidirectional interfaces for system services to optimize performance. The major
features of B-MAC include using clear channel assessment (CCA) and packet backoff for channel arbitration, link layer
acknowledgments for reliability, and low power listening (LPL) for low power communication. B-MAC achieves over 4.5 times
the throughput of the standard S-MAC unicast protocol because B-MAC has lower per-packet processing and effective CCA.

Z-MAC \cite{mac-z} is a hybrid MAC protocol for wireless sensor networks that combines the strengths of TDMA and CSMA while
offsetting their weaknesses. Z-MAC achieves high channel utilization and low-latency under low contention under CSMA and
achieves high channel utilization under high contention and reduces collision among two-hop neighbors at a low cost under
TDMA. Z-MAC has high cost in setup phase. During setup phase, the node would discover neighbors, obtain time slot and
exchange local frame. The two-hop neighbor list is used as input to a time slot assignment algorithm. Z-MAC assigns time
slots to every node in the network and ensures a broadcast schedule where no two nodes within a two-hop communication
neighborhood are assigned to the same slot. The distributed time slot algorithm is called DRAND, which is based on RAND
algorithm in \cite{rand}. Currently the time slot assignment algorithm only supports the static topology. The issues
incurred by dynamic topology is left untouched. In one-hop Mica2 benchmark, the throughput of Z-MAC is larger than B-MAC
when the contention gets intensive because Z-MAC runs on TDMA when high contention. In one-hop ns2 benchmark, the throughput
of Z-MAC is larger than B-MAC when the contention gets intensive.

The packet delivery latency due to periodic sleep schedule is addressed in \cite{mac-d}. It is desirable to maintain the
energy efficiency from duty cycling and reduce the sleep latency in the same time. The paper designs algorithm to assign
time slots to minimize the maximum delay between sensor nodes that can communicate in an arbitrary pattern.

\section{L-MAC Protocol Design} \label{pd}

As we mention in previous section, the MAC protocol design in
wireless sensor network should be given special consideration on the
energy efficiency. To best understand how we design our L-MAC
protocol, we describe first three main energy wastes in MAC layer.
Avoiding these energy wastes in protocol should be always kept in
mind when we design MAC protocols for wireless sensor network.

\begin{itemize}
  \item \textbf{Collision}: Collision is a major resource of energy waste.
  Collision directly leads to packet loss and retransmission, which
  obvious wastes energy. The ability to avoid collision almost
  determines the performance of MAC, especially the energy
  efficiency.
  \item \textbf{Overhearing}: The signal transmitted by the source can
  usually cover a region consisting of several nodes. If it is not
  broadcast, then some nodes may hear the signal that are not for
  him. The results is that these nodes will waste energy in
  receiving and decoding the packets though they are useless.
  \item \textbf{Idle Listening}: It has been identified in previous
  literature, the sensor spends almost the same power in idle as it does
  in receiving or transmitting modes.
\end{itemize}

In L-MAC, we implements several techniques to solve the three energy
waste problems described above. To avoid the collision, we divide
the whole network into to small blocks in geography, assign time
slot to them to avoid collision in block-to-block communication. To
avoid the overhearing problem, nodes go to sleep mode once it hears
some on-going communications between other nodes. To avoid idle
listening, we let sensor nodes go to sleep when they are not
transmitting or receiving packets. In the following, we describe
L-MAC protocol in four parts.

\subsection{Network Division}

\begin{figure} [ht]
\centerline{\includegraphics[scale=0.45]{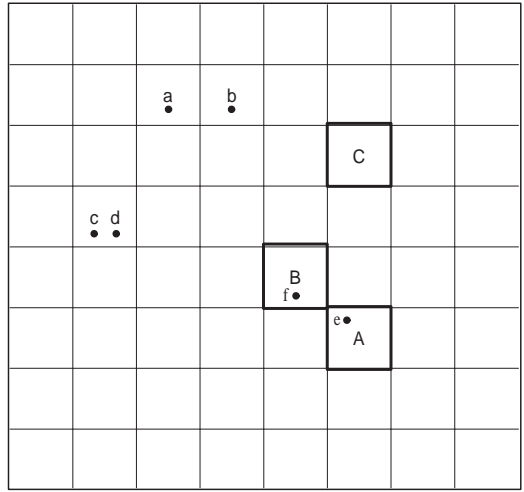}}\caption{Demostration
of Network Division} \label{netdiv}
\end{figure}

In L-MAC, the network is divided into small blocks each of which
contains several sensors. The size and the shape of the block are
application specific. In this work, we assume the network is divided
into square blocks with the side length equal to the one-hop
transmission range of the sensor as shown in Fig.~\ref{netdiv}

Network division is performed at the design phase of the sensor
network application. Since it is application specific, it remains
unchanged unless the requirement of application changes. It will not
affected by new node deployment or node failure, and it will not
change even when nodes move around.

Obviously, any pair of one-hop communicating nodes can either be
within the same block or within two different blocks. Accordingly,
we define the following two concepts: inter-block communication and
intra-block communication.

\begin{itemize}
  \item \textbf{Inter-block communication} - if the node and its one-hop
  communicating neighbor are within different blocks. Obviously, in
  our network division, they must be within adjacent blocks, for an
  example, node $A$ and $B$ in Fig.~\ref{netdiv}.
  \item \textbf{Intra-block communication} - if the node and its one-hop
  communicating neighbor are within the same blocks, for an example,
  node $C$ and $D$ in Fig.~\ref{netdiv}.
\end{itemize}

Note that increasing or decreasing the size of block will affected
the number of inter-block and intra-block communications.

\subsection{Slot Schedule Setup}

In this section, we will show how we introduce the TDMA mechanism
into the L-MAC protocol. We describe two important concepts and then
discuss the algorithm for time slot assignment in L-MAC.

\subsubsection{Inter-/Intra- Block Slot}
L-MAC introduces TDMA mechanism into the protocol. Each
block will be assigned two time slots, one for inter-block
communication and the other for intra-block communication. We call
the time slot \emph{inter-block slot} if it is assigned to the block
for its inter-block communication, and \textit{intra-block slot} if
it is assigned to the block for its intra-block communication. We
restrict that any inter-block communication should be completed in
inter-block slot, and any intra-block communication should be
completed in intra-block slot. Note that, none of inter-block slot
and intra-block slot are overlapped.

We give an example here to clarify how the above mechanism works.
Let us revisit Fig.~\ref{netdiv}, and consider a block $\mathbf{A}$.
If two nodes in block $\mathbf{A}$ want to conduct a direct
transmission\footnote{ Direct transmission means they are within
one-hop distance from each other}, they have to wait until the
intra-block slot of $\mathbf{A}$. If node $e$ in block $\mathbf{A}$
 and node $f$ in block $\mathbf{B}$ want to conduct a direct
 transmission, node $e$ has to postpone the transmission until the
 beginning of inter-block slot of $\mathbf{B}$ if node $e$ is the
 sender and vice versa.

\subsubsection{Time Slot Reuse}
A very important reason that we divide the network into blocks
geographically is that blocks far away from each other can share the
same time slot (inter-block slot or intra-block slot or both)
without interfering each other. For an example, in our configuration
of network as shown in Fig.~\ref{netdiv}, block $\mathbf{A}$ and
block $\mathbf{C}$ can share the same inter-block slot and
intra-block slot without interfering each other as any node in block
$\mathbf{A}$ is at least two-hop distance from any node in block
$\mathbf{C}$ and vice versa.

\vspace{0.2in}

 Although there may be different criteria to design the
algorithm that assign  inter-block slot and intra-block slot to each
block in the network,  it is straightforward that one reasonable
criterion is to design an algorithm that needs the least total
number of time slots for the whole network as this reduce the delay
of transmission. We will not focus on any specific algorithm to
achieve this goal in order to leave our protocol more flexibility.
However, we show an example of time slot assignment that is
implemented  in our work  to reveal the basic idea.

\subsubsection{Inter-block Slot Assignment} 
\begin{figure} [ht]
\centerline{\includegraphics[scale=0.45]{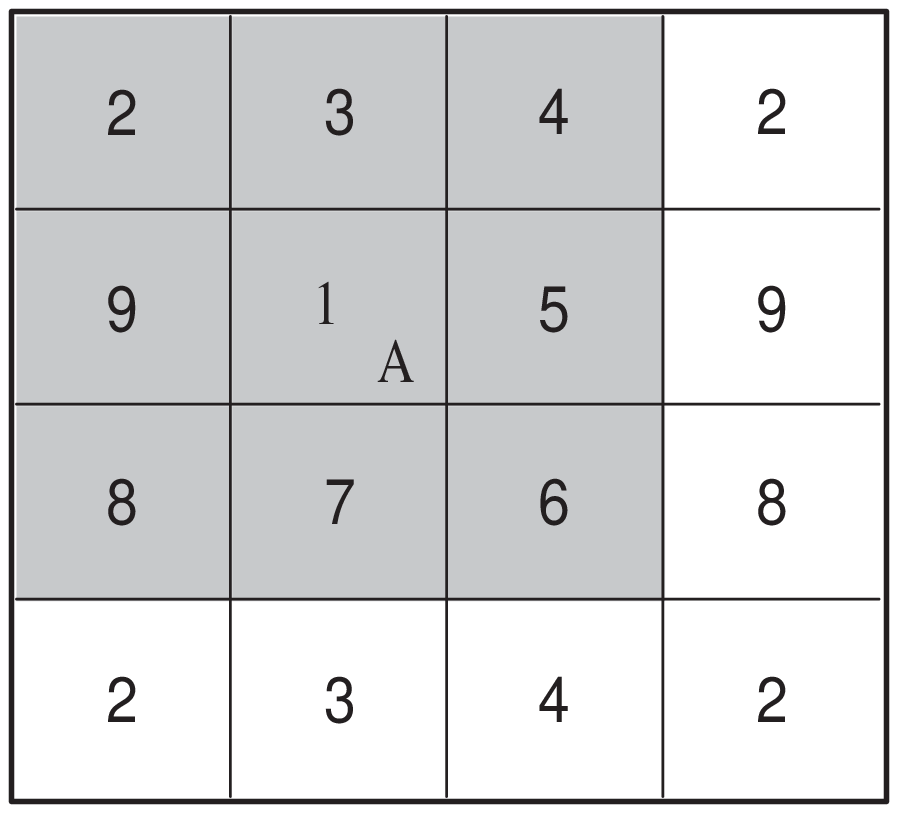}}\caption{Example of
inter-block slot assignment. The number in the block is the
inter-block slot assigned to this block.} \label{inter}
\end{figure}

We depict 16 blocks in
Fig.~\ref{netdiv} separately in Fig.~\ref{inter}. If we assign time
slot 1 to block $\mathbf{A}$ as inter-block slot, then to eliminate
the interference, we have to assign different slots to
$\mathbf{A}$'s surrounding 8 blocks. As shown in Fig.~\ref{inter}, a
total of 9 time slots are used for the inter-block slot assignment
of these 9 blocks. Since in our configuration, the side length of
each block is equal to the one-hop transmission range, we can reuse
these 9 time slots in the blocks that are at least two-hop away, as
illustrated in Fig.~\ref{inter}. Then the whole network will just
need 9 time slot inter-block slot assignment.

\subsubsection{Intra-block Slot Assignment} 
\begin{figure} [ht]
\centerline{\includegraphics[scale=0.45]{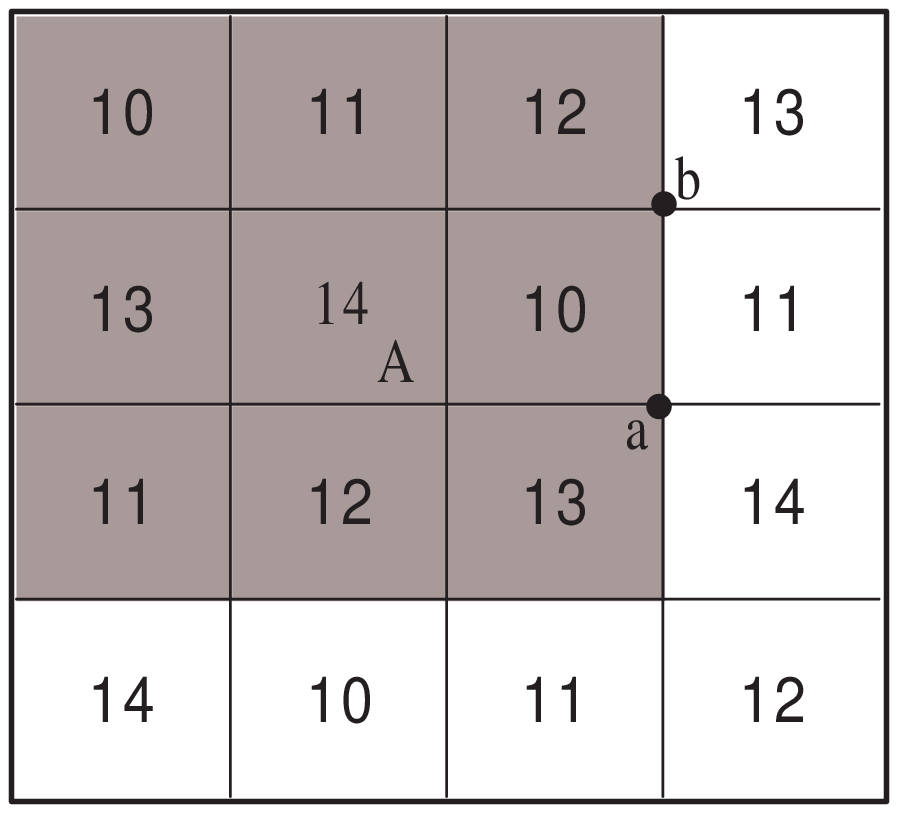}}\caption{Example of
intra-block slot assignment. The number in the block is the
intra-block slot assigned to this block.} \label{intra}
\end{figure}

In intra-block slot,
node can only communicate with nodes in the same block. Therefore,
the intra-block communication will not result in the interference to
any other communication that is at least one-hop away. For our
simple configuration of network division, a intra-block slot
assignment can be given as in Fig.~\ref{intra}.

As we can see in Fig.~\ref{intra}, the node in any block is at least
one-hop away from the node in the closest block which share the same
intra-block slot. The only possible collision occurs, for an
example, when node $a$ and $b$ are both in activity. And with reuse
of intra-block slot, we use only 6 time slot for the intra-block
slot assignment of the whole network.

It should be noted that though we use time slots 1-9 for inter-block
slot and time slots 10-14 for intra-block slot in the above example,
we have no restriction on the assignment of inter- or intra- block
slot. The only restriction that should be kept in mind is that any
time slot assignment should avoid the interference resulted from the
reuse of time slot. Actually, another assignment algorithm developed
from the above example only needs just 9 time slots to complete the
inter-/intra-block slot assignment for the whole network.

\subsection{Inter-block Communication}

In this section, we describe how L-MAC protocol achieves energy
efficiency in inter-block communication.

\subsubsection{First-In First-Receive (FIFR) Rule}
At the beginning of the inter-block slot of block $\mathbf{A}$, all
nodes in will wake up and waiting for the incoming packet. To avoid
collision, we do not allow two nodes to receive packets at the same
inter-block slot if the packets come from different senders. Thus,
to decide which node should keep awake, we design the First-In
First-Receive (FIFR) Rule as below.

\textit{\textbf{First-In First-Receive (FIFR) Rule}}: The node
hearing the packet destined for itself at the earliest time will
keep alive to receive the packet. It will response to that incoming
packet so that every other node within the same block will be
informed and they will go to sleep mode.

It is possible that there is no incoming packets for a block in the
whole inter-block slot. Considering this situation, we set a
threshold of $\theta$ where $0 < \theta <1$ that when there is still
no incoming packets after $\theta$ of one inter-block slot elapses,
all nodes in the block will go to sleep.

\subsection{Intra-block Communication}

Since the number of nodes in one block is not large as we
divide the network, we simply use CSMA/CA mechanism for
the intra-block communications. All nodes will wake up in
intra-block slot. They will possibly transmit packets to each
other or exchange neighborhood information.

\section{Simulation Results} \label{sr}

To evaluate the performance of L-MAC, we implement L-MAC and run Z-MAC, and S-MAC in NS-2 (\cite{ns2}) respectively.
We randomly distribute nodes in 800 × 800 m2
area. We
assume that the nodes know their own positions and we set
the transmission range of 250m. The side length of each block
is set to 200m. The packet size is 512 bytes unless otherwise
specified. The simulation runs for 500 seconds. The metrics
we used to compare different MAC schemes are: (1)energy
consumption. We divide sensor node operation as four modes:
sleep, active (idle), transmit, and receive. We calculate energy
consumption in each mode and add them together. (2)end-end
delay.

\section{Conclusion and Future Works} \label{cl}

In this paper, we presented a new energy and delay efficient
MAC protocol, which features a simple, flexible implementation and robust to network changes and node failure. The
beauty of L-MAC comes from its novel scheme and optimal
time slot assignment algorithm. In L-MAC, time slots are
assigned to each block and each block operates within two
time slots: inter slot and intra slot. With optimal time slot
assignment algorithm, L-MAC can achieves impressive energy
and delay efficiency. L-MAC also successfully solve the
issue of node movement and node failure. Simulation results
demonstrate that L-MAC has a better energy efficiency than
SMAC and L-MAC (with medium to high traffic), and that
L-MAC always presents better delay properties.
Future work may include parameter analysis and be extended to implement L-MAC on sensor motes. We expect to
compare and test our L-MAC protocol in a more comprehensive and realistic way.

\end{document}